\documentclass[dvips]{acta}
\usepackage{supertabular,lscape,epsfig,rotating}
\usepackage{amssymb}
\usepackage{amsmath}
\DeclareSymbolFont{ppa}{OT1}{ppl}{m}{it}
\DeclareMathSymbol{\vv}{\mathalpha}{ppa}{'166}

\newfont{\hb}{rphvb at 10pt}
\newfont{\hbo}{rphvbo at 10pt}
\newfont{\bitt}{rptmbi at 12pt}
\newfont{\bits}{rptmbi at 11pt}

\SetPages{199}{216}

\SetVol{61}{2011}

\begin{document}

\newcommand{\TabApp}[2]{\begin{center}\parbox[t]{#1}{\centerline{
  {\bf Appendix}}
  \vskip2mm
  \centerline{\small {\spaceskip 2pt plus 1pt minus 1pt T a b l e}
  \refstepcounter{table}\thetable}
  \vskip2mm
  \centerline{\footnotesize #2}}
  \vskip3mm
\end{center}}

\newcommand{\TabCapp}[2]{\begin{center}\parbox[t]{#1}{\centerline{
  \small {\spaceskip 2pt plus 1pt minus 1pt T a b l e}
  \refstepcounter{table}\thetable}
  \vskip2mm
  \centerline{\footnotesize #2}}
  \vskip3mm
\end{center}}

\newcommand{\TTabCap}[3]{\begin{center}\parbox[t]{#1}{\centerline{
  \small {\spaceskip 2pt plus 1pt minus 1pt T a b l e}
  \refstepcounter{table}\thetable}
  \vskip2mm
  \centerline{\footnotesize #2}
  \centerline{\footnotesize #3}}
  \vskip1mm
\end{center}}

\newcommand{\MakeTableApp}[4]{\begin{table}[p]\TabApp{#2}{#3}
  \begin{center} \TableFont \begin{tabular}{#1} #4 
  \end{tabular}\end{center}\end{table}}

\newcommand{\MakeTableSepp}[4]{\begin{table}[p]\TabCapp{#2}{#3}
  \begin{center} \TableFont \begin{tabular}{#1} #4 
  \end{tabular}\end{center}\end{table}}

\newcommand{\MakeTableee}[4]{\begin{table}[htb]\TabCapp{#2}{#3}
  \begin{center} \TableFont \begin{tabular}{#1} #4
  \end{tabular}\end{center}\end{table}}

\newcommand{\MakeTablee}[5]{\begin{table}[htb]\TTabCap{#2}{#3}{#4}
  \begin{center} \TableFont \begin{tabular}{#1} #5 
  \end{tabular}\end{center}\end{table}}

\newfont{\bb}{ptmbi8t at 12pt}
\newfont{\bbb}{cmbxti10}
\newfont{\bbbb}{cmbxti10 at 9pt}
\newcommand{\uprule}{\rule{0pt}{2.5ex}}
\newcommand{\douprule}{\rule[-2ex]{0pt}{4.5ex}}
\newcommand{\dorule}{\rule[-2ex]{0pt}{2ex}}
\def\thefootnote{\fnsymbol{footnote}}
\begin{Titlepage}
\Title{The Optical Gravitational Lensing Experiment.\\
High Proper Motion Stars in the OGLE-III Data\\ for Magellanic Clouds Fields
\footnote{Based on observations obtained with the 1.3~m Warsaw telescope at 
the Las Campanas Observatory of the Carnegie Institution for Science.}}

\Author{R.~~P~o~l~e~s~k~i$^1$,~~ I.~~S~o~s~z~y~ñ~s~k~i$^1$,~~ A.~~U~d~a~l~s~k~i$^1$,\\ 
M.\,K.~~S~z~y~m~a~ñ~s~k~i$^1$,~~ M.~~K~u~b~i~a~k$^1$,~~
G.~~P~i~e~t~r~z~y~ñ~s~k~i$^{1,2}$,\\
£.~~W~y~r~z~y~k~o~w~s~k~i$^{1,3}$~~ and~~ K.~~U~l~a~c~z~y~k$^1$}
{$^1$ Warsaw University Observatory, Al. Ujazdowskie 4, 00-478 Warszawa, Poland\\
e-mail: (rpoleski,soszynsk,udalski,msz,mk,pietrzyn,wyrzykow,kulaczyk)@astrouw.edu.pl\\
$^2$ Universidad de Concepción, Departamento de Fisica, Casilla 160-C, Concepción, Chile\\
$^3$ Institute of Astronomy, University of Cambridge, Madingley Road, Cambridge CB3~0HA,~UK
}

\Received{September 13, 2011}
\end{Titlepage}

\Abstract{
We present the results of a search for High Proper Motion (HPM) stars,
\ie the ones with $\mu>100$~mas/yr, in the direction to the Magellanic Clouds. 
This sky area was not examined in detail as the high stellar density
hampers efforts in performing high-quality astrometry.

Altogether 549 HPM stars were found with median uncertainties of proper
motions per coordinate equal to 0.5~mas/yr. The fastest HPM star has the
proper motion of $722.19\pm0.74$~mas/yr. For the majority of objects (70\%)
parallaxes were also measured. The highest value found is
$\pi=91.3\pm1.6$~mas. The parallaxes were used to estimate absolute
magnitudes which enriched with color information show that 21 of HPM stars
are white dwarfs. Other 23 candidate white dwarfs were selected of HPM
stars with no measurable parallaxes using color--magnitude diagram. The
search for common proper motion binaries revealed 27 such pairs in the
catalog. The completeness of the catalog is estimated to be $>80\%$ and it
is slightly higher than for previous catalogs in the direction to the
Magellanic Clouds.}{Astrometry -- Catalogs -- Stars: kinematics and
dynamics -- binaries: visual -- white dwarfs}

\Section{Introduction}
\hglue-9pt Previous surveys searching for High Proper Motion (HPM) stars 
typically avoided dense stellar regions of the Magellanic
Clouds (MCs; \eg Finch \etal 2007 as a part of SuperCOSMOS-RECONS
survey). First, high stellar density hampers HPM star identification
especially in two-epoch surveys. Second, the probability of blending of a
HPM star with a background object which lowers the observed proper motion
of the light centroid of both objects is high (Koz³owski \etal 2006). The
SPM4 survey (Girard \etal 2011) covers the area of MCs, but it uses only
two epochs and suffers from problems with star identification. To search
for HPM objects Alcock \etal (2001) and Soszyñski \etal (2002) used,
respectively, the MACHO and the second phase of the Optical Gravitational
Lensing Experiment (OGLE-II) surveys data, which both aimed at searching
for microlensing events.  In this paper we extend the study of Soszyñski
\etal (2002) using the data collected during the third phase of the OGLE
survey (OGLE-III), which covered larger sky area (54 \vs 8.3~deg$^2$),
longer time baseline (8 \vs 4 years) and had better pixel scale (0.26 \vs
0.46~arsec/pixel). Since the OGLE-III was a photometric survey, the
observing strategy was not optimized for astrometry. However, the large
number of epochs taken at good seeing conditions allowed us to measure
proper motions and parallaxes at milliarcsecond level with milliarcsecond
precision.

In Section~2 we describe observations and data reduction which was
performed before we started this research. Sections 3 and 4 describe
calculation of PM and selection of HPM objects. In Section~5 we describe
the catalog, asses its completeness, compare with other catalogs, analyze
color--absolute magnitude and color--magnitude diagrams (CMDs) as well as
list common proper motion (CPM) binaries.

\vspace*{-9pt}
\Section{Observations and Initial Data Reduction}
\vspace*{-5pt}
The OGLE-III observations were carried out between 2001 and 2009 with 1.3-m
Warsaw telescope located at Las Campanas Observatory, Chile. The
observatory is operated by the Carnegie Institution for Science. The only
one instrument attached to the Warsaw telescope at that time was the
``second generation'' camera\footnote{Note that Anderson \etal (2006) found
distortion corrections accurate to $\approx7$~mas for individual frames
taken with WFI camera on 2.2-m ESO telescope and $\approx100$~mas changes
caused by the manipulation of the camera.}. It consisted of eight
$2048\times4096$ pixel CCD detectors with 15 $\mu{\rm m}$ pixels which gave
0.26 arcsec/pixel scale. The total field of view was $35\times35.5$~arcmin.
Around 90\% of the images of the MCs were taken with the {\it I} filter and
the remaining with the {\it V} filter. We analyzed only {\it I}-band images
number of which varied between 385 and 637 for the Large Magellanic Cloud
(LMC) fields and between 583 and 762 for the Small Magellanic Cloud (SMC)
fields with the exception of the field SMC128 for which 1228 epochs were
secured. The time baseline was 4.5~yr for the field SMC140 (covering the
center of the globular cluster 47~Tuc) and between 7.5~yr and 7.9~yr for
the remaining fields. The details of the instrumentation setup were given
by Udalski (2003).

\begin{figure}[htb]
\centerline{\includegraphics[angle=270,width=.95\textwidth]{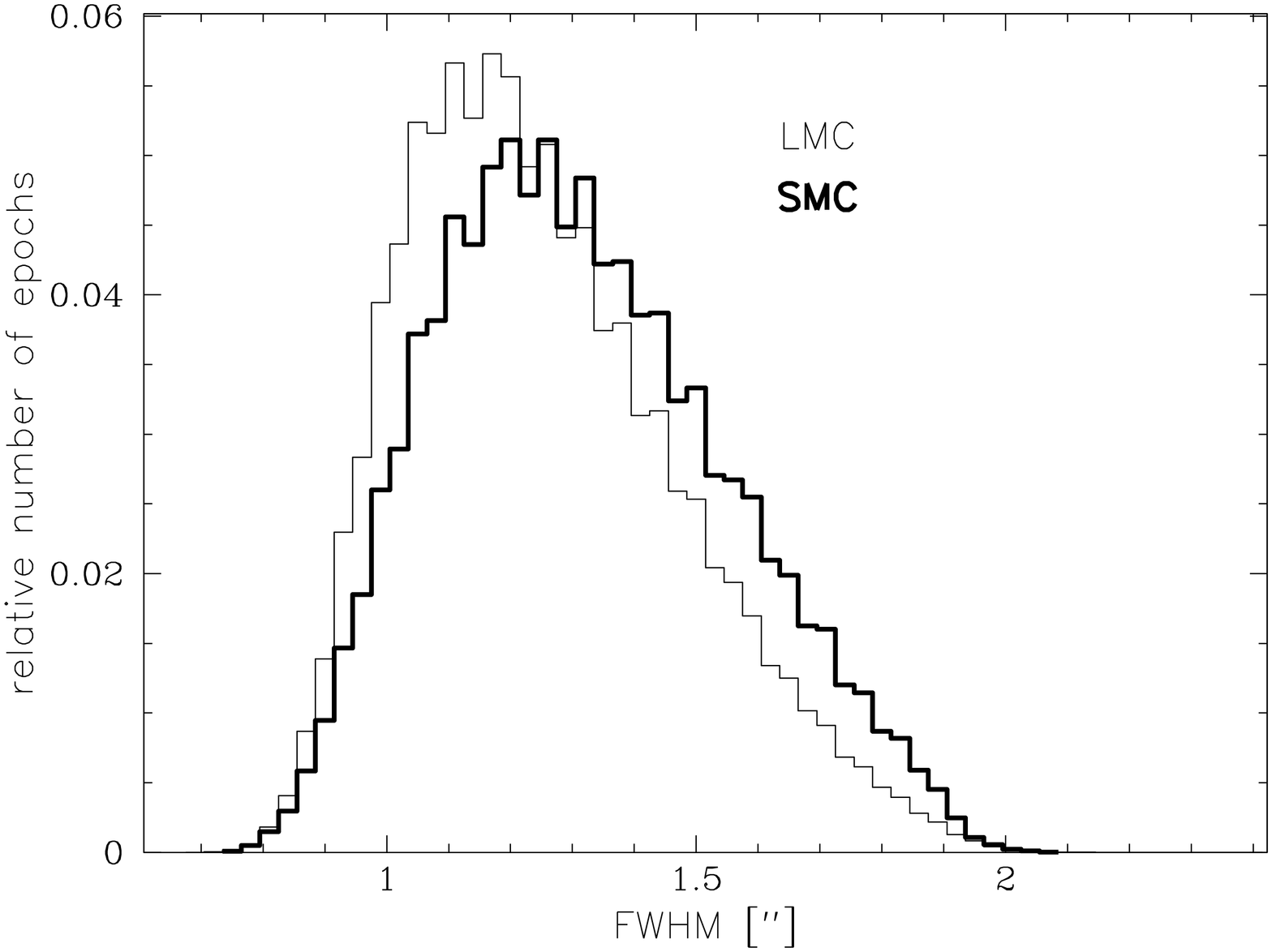}}
\FigCap{Relative number of epochs taken with given FWHM 
of stellar profiles for LMC (thick line) and SMC (thin line)
separately. The bin size is $0\zdot\arcs03$.}
\end{figure}

Hardly ever a single MCs field was observed more than once per night during
the OGLE-III survey. Centers of the observed fields were fixed and the only
shifts resulted from the telescope pointing errors. Fig.~1 shows the
relative number of exposures for a range of seeing for the LMC and SMC
fields separately. The median seeing was $1\zdot\arcs2$ and one can see a
slightly worse seeing for SMC caused by the higher airmasses at which the
images were taken. The minimum airmasses of the LMC and SMC fields during
observations were 1.26 and 1.34, respectively. The images were taken at
airmasses up to 2. The standard OGLE-III photometry was obtained with the
Difference Image Analysis (DIA) method (Alard and Lupton 1998, Alard 2000,
Wo¼niak 2000). DIA requires reference images which were constructed using
up to 30 best images of a given field (Udalski \etal 2008a). The direct
transformation from the pixel coordinates to the equatorial coordinates is
available only for the reference images. The transformation was based on
the Two Micron All Sky Survey (Skurstkie \etal 2006) catalog and gave
$0\zdot\arcs12$, {\it rms} per coordinate when compared to the Third US
Naval Observatory CCD Astrograph Catalog (Zacharias \etal 2010). The
photometric maps of the MCs based on the OGLE-III reference images were
published by Udalski \etal (2008b, 2008c).

The preparation of the reference images began with the selection of the
best images for each subfield (which corresponds to a single CCD chip of
the mosaic) separately. Each subfield was divided into two or eight
subframes (the size was either $2180\times2088$ or $1090\times1044$~pixels)
and analyzed separately. The images used for the reference image
construction were resampled to the grid of the best quality subframe on the
list. Before the resampling was done a cross-match was performed on the
list of bright stars found on the given image and the reference image. For
the MCs fields a typical brightness limit for stars used in the
cross-matching was between 18 and 18.5 mag and depended on seeing, airmass,
etc. Next, to find the transformation between the analyzed and the
reference subframes, the third order polynomial was fitted. To remove
outliers from the fit, 3$\sigma$-clipping was performed and a new set of
polynomial coefficients were found. Then, the analyzed subframe was
resampled to the reference grid using spline method and the polynomial
found above. Finally, all the resampled subframes were co-added in the way
that each pixel was composed of up to 10 pixels from the resampled
subframes. The advantage of the spline resampling is that it conserves the
total flux of each star.

Each individual image was divided into subframes corresponding to the
subframes of the reference image, and each of them was resampled to the
grid of the reference subframe. The procedure of calculating transformation
and resampling was the same as described above. It allows the indirect
calculation of the equatorial coordinates for each epoch separately.
Further reduction for the standard OGLE photometry was performed using the
DIA package. All the transformed images were also reduced using the {\sc
DoPHOT} software (Schechter \etal 1993) and this study is based on these
data only. This additional reduction allows a simple selection of
foreground Galactic objects based on the time series astrometry. The {\sc
DoPHOT} performs analytical model point spread function (PSF) fitting and
was run in the spatially-variable PSF mode. The stellar positions measured
on each image were associated with the stellar object on the reference
image, where matching radius of 1.9 pixel was used. These data were used \eg
to select Galactic variable stars in an on-going research of the OGLE-III
Catalog of Variable Stars (\eg Soszyñski \etal 2009 lists 66 Galactic
RR~Lyr type variables). For some stars (\eg the ones with brightness close
to the saturation limit) the {\sc DoPHOT} photometry is more robust than
the DIA one. The details of the reference image construction and obtaining
the {\sc DoPHOT} photometry were described by Udalski \etal (2008a).

The procedure described above was far from being optimal for astrometric
purposes. We describe their main disadvantages below. First, transformation
of grids was performed on centroids found using one program ({\sc sfind} in
this case; Wo¼niak 2000) and after the transformation was done the second
time centroids were found using another program ({\sc DoPHOT}). The results
of both programs might be different. In the exceptional case we found a
mean difference of 0.1~pixel and the inspection of the image revealed that
even though the seeing was better than average the shape of the PSF was
elongated and the {\sc DoPHOT} treated each star as two separate objects,
which affected measured centroids. Second, the transformation was
calculated using all the bright stars, including HPM objects. Also very red
or very blue objects affected calculated transformation, as the
differential refraction effect was not removed.  Third, the spline
resampling does not conserve relative positions of centroids. This effect
had very slight influence on the measured positions. Fourth, the measured
centroids were cross-referenced with the database records with a constant
radius, so the fastest HPM stars were at different epochs associated with
the different database records. It was also because the reference images
were constructed using the stellar positions from different epochs and HPM
stars have elongated profiles or even the profiles have more than one local
maximum.

With the last of the disadvantages mentioned we could cope, because the raw
{\sc DoPHOT} results were secured and can be queried around a certain
position on the chip. Proper removing of the first two disadvantages was
somewhat more complicated. It would require almost the entire reanalysis of
a huge dataset. Instead of doing this very computer time consuming task we
decided to apply a post-mortem corrections to the measured positions which
reduced these effects. Even though we did not fully remove them our results
gave very precise and reliable proper motions in the dense stellar fields.

\Section{Method of Proper Motion Calculation}
The quantities which affect the position of the star and are important from
astrophysical point of view are the proper motion and the parallax.
Ground-based observations are also affected by the atmospheric refraction
and the annual aberration. Telescope pointing removed aberration for the
field center. For the wide-field imagers one also has to remove
differential aberration caused by the difference in angular separation
between Earth apex and different points of the sky imaged. In the OGLE-III
pipeline differential aberration was removed by the resampling procedure
described above. This also removed the mean atmospheric refraction.
Refraction depends on the spectrum of an object, thus we had to take into
account differential refraction. At time $t$ ($t=0$ for 2000.0) the
observed position of the star ($\alpha$, $\delta$) is given by the
following formulae:
$$\alpha=\alpha_0+\mu_\alpha t+\frac{r\sin p\tan z+\pi\sin\gamma\sin\beta}{\cos\delta}\eqno(1)$$
$$\delta=\delta_0+\mu_\delta t+r\cos p\tan z+\pi\sin\gamma\cos\beta\eqno(2)$$
where $\mu_\alpha$ and $\mu_\delta$ are proper motions along the right
ascension and the declination. Differential refraction coefficient and
parallax are designated $r$ and $\pi$, respectively, $z$ is the zenith
distance, $\alpha_0$ and $\delta_0$ are coordinates for the J2000.0
equinox, $\gamma$ is the angular distance to the Sun, $\beta$ and $p$ are
angles between direction of parallax shift and refractional shift,
respectively, and direction to the North celestial pole.

Kuijken and Rich (2002) showed that the uncertainty with which one measures
the centroid of a star ($\sigma_{\rm PSF}$) depends on the Full Width at
Half Maximum (FWHM) of the stellar profile and the signal to noise (S/N)
ratio of the stellar flux:
$$\sigma_{\rm  PSF}=\frac{0.67\cdot{\rm FWHM}}{\rm S/N}.\eqno(3)$$
We measured the FWHM for each OGLE-III subframe and S/N values were
estimated using uncertainties of brightness ($\sigma_m$~[mag]) and the
standard relation:
$$\sigma_m=\frac{1.086}{\rm S/N}.\eqno(4)$$
The uncertainty $\sigma_{\rm PSF}$ reflects only a contribution from a
finite number of ADU and nonzero seeing. The total uncertainty of the
centroid position also depends on how well one can fit the grid of a given
subframe to the grid of the reference subframe. This factor was taken into
account later.

Determination of proper motions started with dividing the list of stars
from each subfield into subframes in the same manner as in Udalski \etal
(2008a). This was mandatory as each subframe was reduced separately and the
zero point of proper motions may be different. The rest of the calculations
were performed for each subframe independently. For each star the
time-series astrometry in the pixel scale of the reference image and
photometry were obtained from the database of {\sc DoPHOT} results. Pixel
scale was transformed to the equatorial coordinates using transformation
for the reference images. For each star and each epoch separately the
angles $p$, $z$, $\gamma$ and $\beta$ were calculated using ephemerides by
van Flandern and Pulkkinen (1979). The uncertainties of centroid
measurements $\sigma_{\rm PSF}$ were calculated using Eqs.~(3) and (4).

The goal of the following calculations was to find for each epoch and each
subframe corrections for positions of centroids that would be used for
calculation of proper motions in the reference frame of background objects.
By background we assumed 47~Tuc for the field SMC140 and MCs for the
remaining fields\footnote{Some parts of the SMC131, SMC136 and SMC137
fields are also within the tidal radius of 47~Tuc which equals
42\zdot\arcm9 (Kiss \etal 2007).}. See Anderson and King (2003), Girard
\etal (2011), van der Marel \etal (2002) and Piatek \etal (2008) for a
discussion of absolute proper motions of background objects. Other
important quantities are: correction of centroid uncertainties, dispersion
of proper motions of background stars and average $(V-I)$ \vs $r$
relations. Accurate calculation of these quantities is possible only if
bright stars are taken into account. We selected stars brighter than 18~mag
in {\it I}-band and with color information (hereinafter ``good''
stars). For good stars two models were fitted to Eqs.~(1) and (2): one with
non-zero proper motion (free parameters: $\alpha_0$, $\delta_0$,
$\mu_\alpha$, $\mu_\delta$ and $r$) and one with zero proper motions (free
parameters: $\alpha_0$, $\delta_0$ and $r$). We used the $\chi^2$
minimalization procedure and typically had a few hundred equations (twice
the number of centroid measurements for a given star) for five or three
unknowns. The nearest stars with significant parallax might have influenced
the derived corrections thus at each time we found the proper motions, we
excluded the stars with
$\mu=\sqrt{\mu_{\alpha\star}^2+\mu_\delta^2}>20$~mas/yr, where
$\mu_{\alpha\star}=\mu_\alpha\cos\delta$. From the two models we chose the
one with non-zero proper motion if resulting $\chi^2$ was smaller than 0.9
times $\chi^2$ for the zero proper motion model. We note that $\chi^2$
values for bright stars were much larger than unity, as uncertainties of
grid fitting were not included in that iteration. This multiplication
constant was changed to 0.95 in the successive iterations. Using the model
found this way we calculated expected centroid for each epoch and
subtracted it from the observed position. These residua were averaged and
their {\it rms} was found. The average residua were subtracted from the
centroids of the given epoch and results were used in the next iteration.
The centroid uncertainties $\sigma_{\rm PSF}$ were square added to the {\it
rms} of the residua ($\sigma_{\rm GRID}$) giving uncertainties of centroid
positions. These centroids and their uncertainties were used in the next
iteration in which all the models were fitted once more. The proper motions
were used to find $\sigma$-clipped mean proper motion of background
stars. The opposite of this proper motion was included into the corrections
for positions to assure mean proper motion of background stars would be
zero in the next iteration. The {\it rms} of proper motions of background
stars is a measure of our systematic errors and given with subscript {\it
SYS} in the catalog (as opposed to statistical uncertainties derived from
$\chi^2$ fitting which are indicated with subscript {\it STAT}). The
$\chi^2$ values of the final fits were close to the unity for most of the
stars.

All the above calculations intended to find corrections for positions and
their uncertainties, $(V-I)$ \vs $r$ relations as well as systematic
uncertainties in PMs. Finally, we calculated models with non-zero PM for
all stars in a given subframe. We also tried to fit models with parallax
but in $\approx75\%$ of cases the matrix was ill-conditioned. The
calculations were also repeated using only exposures with seeing better
than $4.5~{\rm pix}=1\zdot\arcs17$. This step increased the number of HPM
candidates (see below) by 5\%. Analysis of the stars with $\mu<100$~mas/yr
will be presented in the forthcoming paper.

\begin{figure}[htb]
\centerline{\includegraphics[height=.95\textwidth,angle=270]{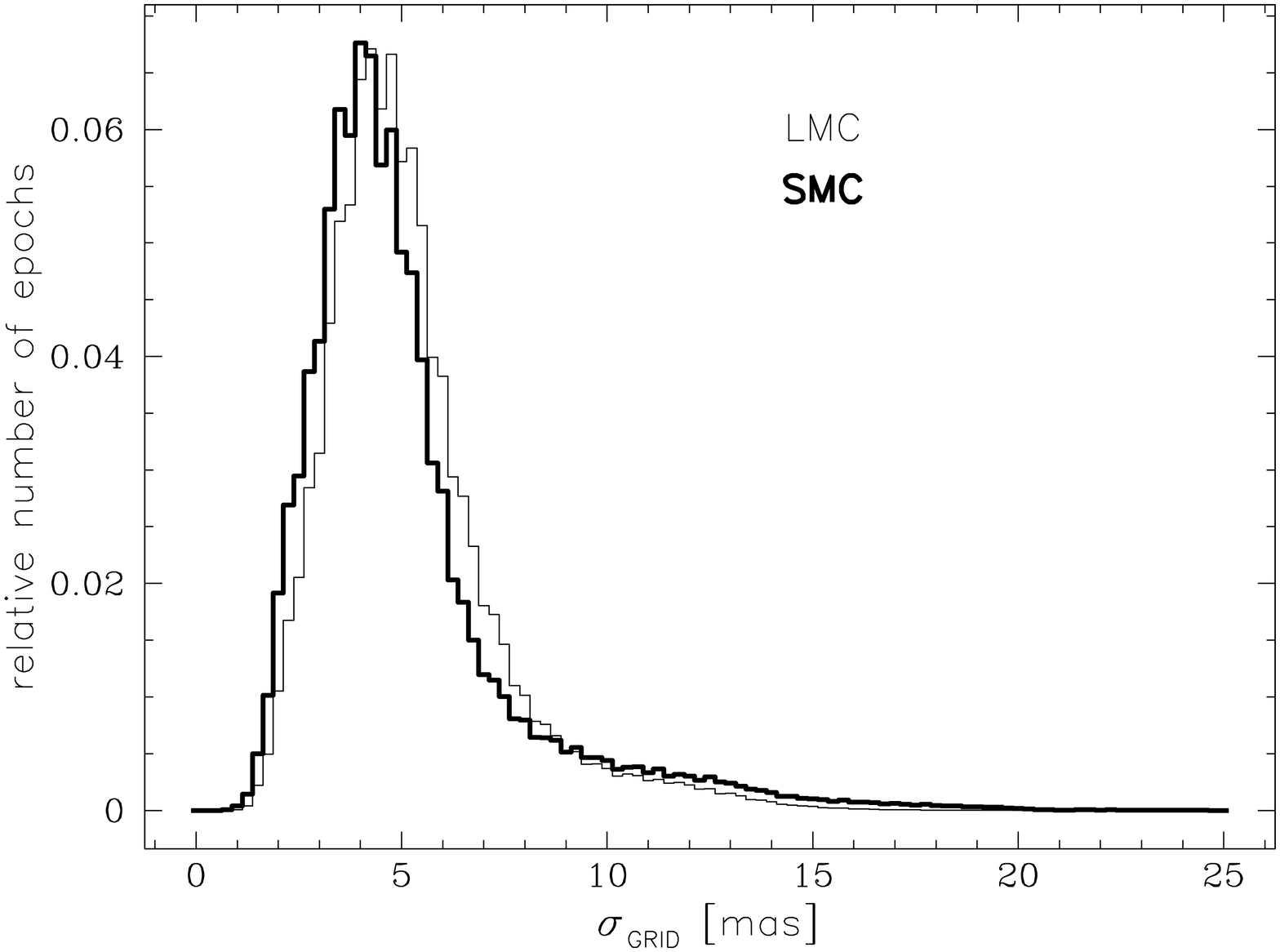}}
\FigCap{Relative histogram of $\sigma_{\rm GRID}$ -- accuracy with 
which we fitted grid of the sub-chip to the grid of the reference sub-chip.
Thin and thick lines correspond to the LMC and the SMC data,
respectively. The bin size is 0.25~mas.}
\end{figure}
\begin{figure}[htb]
\centerline{\includegraphics[width=.95\textwidth]{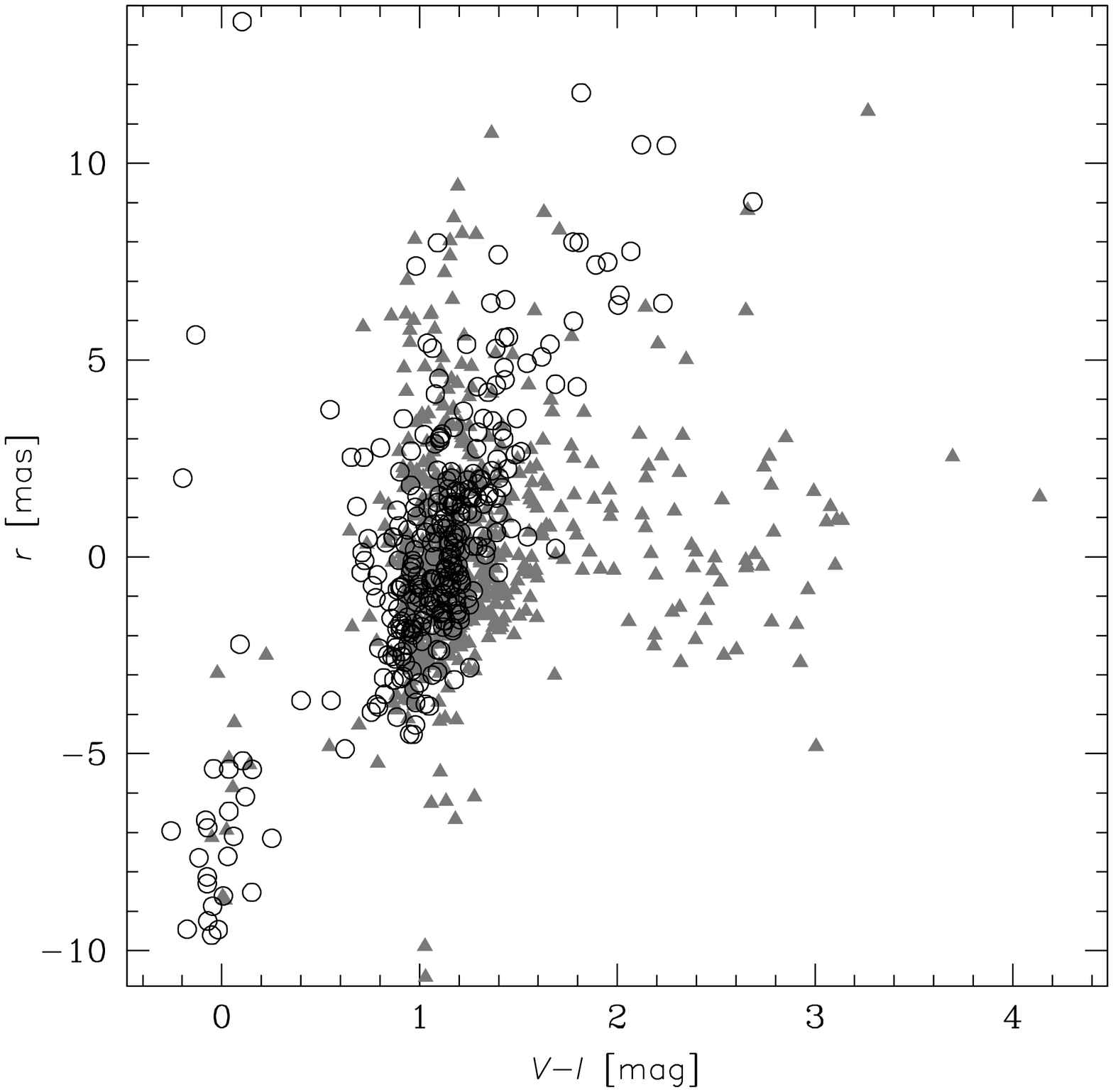}}
\FigCap{Differential refraction coefficient $r$ as a function of $V-I$ 
color for stars in subframes of fields LMC147.3 (black circles) and
LMC112.6 (gray triangles).}
\end{figure}

Fig.~2 presents the relative histogram of epochs with different
$\sigma_{\rm GRID}$ for the LMC and the SMC fields separately. The accuracy
of our relative astrometry is typically in the range $2.5\div 7$~mas which
is close to the best wide-field CCD ground-based images. The smallest
values of $\sigma_{\rm GRID}$ (below 2~mas) and the tail of the
distribution (above 8~mas) were found in the subframes with a small number
of good stars. In these subframes our models were poorly constrained. The
accuracy of the astrometry in the SMC fields was slightly better than in
the LMC fields. Fig.~3 shows the $(V-I)$ \vs $r$ relation for two
subframes.

The uncertainties of the parallax measurements derived from the
least-squares fits ($\sigma_{\pi,{\rm LS}}$) were unreasonably small --
down to 0.12~mas. In order to better assess the quality of our parallax
measurements we have searched the whole dataset of our measurements for
stars that are present in two adjacent fields, in each of them there were
at least 100 epochs and their parallax was greater than 10~mas. For twenty
six such pairs we measured the dispersion of differences in $\pi$ found in
both fields and compared it with $\sigma_{\pi,{\rm LS}}$. This revealed
that $\sigma_{\pi,{\rm LS}}$ are underestimated by 1.5~mas. This component
was added in square to each $\sigma_{\pi,{\rm LS}}$ giving $\sigma_\pi$
claimed in the catalog. Parallax value is given only if $\pi/\sigma_\pi
>3$.

\Section{Selection of HPM Stars}
As HPM stars we defined stars with proper motion greater than 100~mas/yr.
We note that stars with $\mu>126$~mas/yr moved during the OGLE-III time
span enough that they may be present in the OGLE-III photometric maps as
two separate records. Objects with higher proper motion may have even more
records in the database.

We did not use the photometric selection method based on DIA photometry
(Eyer and Wo¼niak 2001). Our tests showed that in the vicinity of bright
stars the DIA package produced a higher number of artifacts resembling the
HPM photometric behavior. The light curves of some HPM stars did not
resemble the expected parabola-like light curves. Both these disadvantages
may originate from the fact that our reference images were composed of
images taken during long period of time.

To select HPM stars we checked all objects with $\mu>95$~mas/yr. Most of
the objects with the uncertainty of the proper motion
$\sigma_\mu>10$~mas/yr had much less than a hundred measurements and proper
motion close to the assumed limit. After inspecting several examples we
decided not to analyze these objects, because their number was very high
and chances of finding additional sound HPM stars were very low.

The list of candidates selected above was compared with the catalogs
presented by Alcock \etal (2001) and Soszyñski \etal (2002). These catalogs
contain altogether 80 unique stars with $\mu>95$~mas/yr. Our list of
candidates contained 76 of them. Two of the stars (MACHO IDs: 2.4668.10 and
5.5613.1633) were overexposed in the OGLE-III photometry. For the object
LMC\_SC8~359715 Soszyñski \etal (2002) gave $\mu=125.6\pm4.0$~mas/yr and
our analysis resulted in $\mu=87.11\pm0.67$~mas/yr. The object is present
in the two adjacent OGLE-III fields and in both of them consistent results
were found. The only one object with $\mu>95$~mas/yr from the lists of
Alcock \etal (2001) and Soszyñski \etal (2002) that was not selected as a
candidate HPM was SMC\_SC10~57257= 206.16886.2221 with $\mu=365.01\pm
0.32$~mas/yr.  Its image in the OGLE-III reference frame is so elongated
that it was categorized as a diffuse object by the {\sc DoPHOT} and thus
not included in the photometric maps. The OGLE-III identification
SMC110.5.999999 was given to that object. We obtained its centroids from
the database and analyzed it in the same manner as the other ones. The
comparison with the MACHO and OGLE-II catalogs revealed our list of
candidates was complete in $76/77\approx99\%$.

For some HPM stars we had more than one record in the candidate list from
the given field, thus we removed double records from the same field of the
same star paying attention not to remove one component of a CPM binary.
For each object we scanned the database to find centroids not associated to
the object, but localized in a close proximity. This way we increased the
number of points used for fitting. Later on, we examined the centroids by
eye to remove some of them \eg measurements with bad seeing, if a HPM was
close to another star. The final fit was performed on data acquired this
way with corrections described in Section~3. Since the OGLE-III neighboring
reference images are overlapping, some objects are present on two, three or
even four adjacent fields. For objects selected above we checked if they
are present in adjacent fields. If the number of measurements in the second
field was significant compared to the number of measurements in the first
field, we performed one more fitting. The model contained separate
positions for the J2000.0 equinox ($\alpha_{0,1}$, $\delta_{0,1}$ and
$\alpha_{0,2}$, $\delta_{0,2}$) and the refraction coefficients ($r_1$ and
$r_2$) for each field. The values of $\mu_\alpha$, $\mu_\delta$ and $\pi$
were kept the same for both fields. This way the number of free parameters
increased from six to nine and the number of equations increased as
well. The objects with unusually high uncertainties of parameters or large
$\chi^2$ values were inspected manually. After calculating the final
models, we removed all the HPM candidates with $\mu< 100$~mas/yr leaving
549 objects.

\Section{Catalog}
The catalog of HPM stars contains altogether 551 objects. Two stars have
$\mu{<}100$~mas/yr but were found to be in CPM binaries with HPM objects. The
catalog is available for the astronomical community only in electronic form
{\it via} FTP site:

\vspace*{-6pt}
\centerline{\it ftp://ftp.astrouw.edu.pl/ogle/ogle3/pm/hpm\_mcs/}
\vskip3pt

For each object we provide the OGLE-III identifier from Udalski \etal
(2008b, 2008c), equatorial coordinates for J2000.0 equinox from our model
fits, proper motion with statistical and systematical uncertainties given
separately, parallax, {\it I}-band brightness, $V-I$ color and luminosity
class (21 WDs, 23 WD candidates and one subdwarf, see Section~5.3). Table~1
shows the exemplary part of the main catalog file. Cross-identifications
with MACHO, OGLE-II and SPM4 catalogs are given in a separate file. The
brightness estimates differ from those in Udalski \etal (2008b, 2008c),
firstly, because they were corrected for the differences in transmission
between standard and the OGLE-III filters (Szymañski \etal 2011).
Secondly, because a more detailed cross-match between {\it V}- and {\it
I}-band images was performed. For each object a finding chart is also
provided with indicated epoch to which it corresponds. We have analyzed the
time-series photometry of all cataloged objects and found periodic
variability for four of them.  The amplitudes are a few hundredths of
magnitude and phased light curves show sinusoid-like variability. These
together with other comments for a few stars are given in {\sl remarks.txt}
file distributed along with the catalog.

\begin{landscape}
\SetTableFont{\scriptsize}
\MakeTable{lrrrrrrrrrrrrrrrl}{22.5cm}{Exemplary part of the main catalog file}
{
\hline
\multicolumn{1}{c}{OGLE-III ID}\uprule &
\multicolumn{1}{c}{R.A.} & 
\multicolumn{1}{c}{Dec.} & 
\multicolumn{1}{l}{\begin{rotate}{-70}$\mu_{\alpha\star}$\end{rotate}} &
\multicolumn{1}{l}{\begin{rotate}{-70}$\sigma_{\mu_{\alpha\star},STAT}$\end{rotate}} &
\multicolumn{1}{l}{\begin{rotate}{-70}$\sigma_{\mu_{\alpha\star},SYS}$\end{rotate}} &
\multicolumn{1}{l}{\begin{rotate}{-70}$\mu_\delta$\end{rotate}} &
\multicolumn{1}{l}{\begin{rotate}{-70}$\sigma_{\mu_\delta,STAT}$\end{rotate}} &
\multicolumn{1}{l}{\begin{rotate}{-70}$\sigma_{\mu_\delta,SYS}$\end{rotate}} &
\multicolumn{1}{l}{\begin{rotate}{-70}$\mu=\sqrt{\mu_{\alpha\star}^2+\mu_\delta^2}$\end{rotate}} &
\multicolumn{1}{c}{$\pi$} &
\multicolumn{1}{c}{$\sigma_\pi$} &
\multicolumn{1}{c}{{\it I}} &
\multicolumn{1}{c}{$\sigma_I$} &
\multicolumn{1}{c}{$V-I$} &
\multicolumn{1}{c}{$\sigma_{(V-I)}$} &
\multicolumn{1}{c}{lumin.} 
\\
\\
\\
\\
\dorule &
\multicolumn{1}{c}{J2000.0} &
\multicolumn{1}{c}{J2000.0} &
\multicolumn{1}{l}{\begin{rotate}{-70}$\rm [mas/yr]$\end{rotate}} &
\multicolumn{1}{l}{\begin{rotate}{-70}$\rm [mas/yr]$\end{rotate}} &
\multicolumn{1}{l}{\begin{rotate}{-70}$\rm [mas/yr]$\end{rotate}} &
\multicolumn{1}{l}{\begin{rotate}{-70}$\rm [mas/yr]$\end{rotate}} &
\multicolumn{1}{l}{\begin{rotate}{-70}$\rm [mas/yr]$\end{rotate}} &
\multicolumn{1}{l}{\begin{rotate}{-70}$\rm [mas/yr]$\end{rotate}} &
\multicolumn{1}{l}{\begin{rotate}{-70}$\rm [mas/yr]$\end{rotate}} &
\multicolumn{1}{c}{$\rm [mas]$} &
\multicolumn{1}{c}{$\rm [mas]$} &
\multicolumn{1}{c}{$\rm [mag]$} &
\multicolumn{1}{c}{$\rm [mag]$} &
\multicolumn{1}{c}{$\rm [mag]$} &
\multicolumn{1}{c}{$\rm [mag]$} &
\multicolumn{1}{c}{class} 
\\
\\
\\
\\
\hline
\uprule
\ldots\\
LMC103.7.10025 & 5\uph17\upm05\zdot\ups90 & $-69\arcd59\arcm06\zdot\arcs0$ & $82.35$  & 0.13 & 0.24 & $221.01$  & 0.13 & 0.25 & 235.85 & 5.03  & 1.57 & 16.636 & 0.012 & 1.197 & 0.020 &  \\ 
LMC103.6.78728 & 5\uph17\upm26\zdot\ups00 & $-69\arcd41\arcm21\zdot\arcs1$ & $92.92$  & 0.32 & 0.29 & $149.81$  & 0.31 & 0.36 & 176.29 & 22.30 & 1.67 & 15.088 & 0.010 & 2.392 & 0.017 &  \\ 
LMC105.6.34014 & 5\uph17\upm57\zdot\ups55 & $-70\arcd54\arcm38\zdot\arcs9$ & $38.86$  & 0.25 & 0.24 & $117.62$  & 0.25 & 0.19 & 123.87 & 9.06  & 1.73 & 17.727 & 0.024 & 3.143 & 0.146 &  \\ 
LMC102.7.22769 & 5\uph18\upm51\zdot\ups10 & $-68\arcd11\arcm17\zdot\arcs7$ & $-4.67$  & 1.65 & 0.32 & $115.80$  & 1.65 & 0.35 & 115.89 & 0.00  & 0.00 & 20.033 & 0.168 & 0.674 & 0.206 & WDcand \\ 
LMC102.7.22886 & 5\uph18\upm54\zdot\ups98 & $-68\arcd09\arcm48\zdot\arcs3$ & $0.73$   & 1.44 & 0.32 & $121.70$  & 1.44 & 0.35 & 121.70 & 0.00  & 0.00 & 19.858 & 0.147 & 0.448 & 0.169 & WDcand \\ 
LMC103.6.38675 & 5\uph18\upm58\zdot\ups18 & $-69\arcd47\arcm47\zdot\arcs5$ & $68.30$  & 0.14 & 0.31 & $105.32$  & 0.13 & 0.36 & 125.53 & 11.35 & 1.57 & 13.497 & 0.029 & 2.120 & 0.030 &  \\ 
LMC103.3.2     & 5\uph19\upm08\zdot\ups19 & $-69\arcd48\arcm13\zdot\arcs8$ & $69.25$  & 0.12 & 0.26 & $104.91$  & 0.12 & 0.25 & 125.70 & 12.39 & 1.55 & 14.147 & 0.061 & 2.559 & 0.063 &  \\ 
LMC104.4.296   & 5\uph19\upm42\zdot\ups85 & $-70\arcd14\arcm01\zdot\arcs4$ & $-72.28$ & 0.22 & 0.24 & $287.16$  & 0.22 & 0.30 & 296.12 & 24.72 & 1.68 & 17.893 & 0.028 & 0.881 & 0.040 & WD \\ 
LMC101.3.36262 & 5\uph19\upm57\zdot\ups25 & $-68\arcd33\arcm04\zdot\arcs5$ & $-38.19$ & 0.11 & 0.26 & $-148.13$ & 0.11 & 0.33 & 152.97 & 7.04  & 1.54 & 15.871 & 0.009 & 2.195 & 0.020 &  \\ 
LMC104.1.32254 & 5\uph20\upm07\zdot\ups20 & $-70\arcd35\arcm31\zdot\arcs7$ & $-42.97$ & 0.14 & 0.24 & $234.53$  & 0.14 & 0.27 & 238.43 & 31.14 & 1.57 & 15.823 & 0.008 & 2.724 & 0.026 &  \\ 
LMC101.1.58593 & 5\uph20\upm10\zdot\ups21 & $-68\arcd52\arcm44\zdot\arcs0$ & $16.11$  & 1.90 & 0.47 & $105.80$  & 1.88 & 0.36 & 107.02 & 0.00  & 0.00 & 20.288 & 0.201 & 0.686 & 0.261 & WDcand \\ 
LMC102.2.30    & 5\uph20\upm14\zdot\ups98 & $-68\arcd10\arcm07\zdot\arcs9$ & $-71.53$ & 0.18 & 0.36 & $121.56$  & 0.18 & 0.34 & 141.04 & 0.00  & 0.00 & 14.859 & 0.014 & 1.929 & 0.021 &  \\ 
LMC102.2.37    & 5\uph20\upm15\zdot\ups12 & $-68\arcd10\arcm08\zdot\arcs1$ & $-71.33$ & 0.18 & 0.36 & $122.67$  & 0.18 & 0.34 & 141.90 & 4.99  & 1.60 & 14.955 & 0.012 & 2.093 & 0.018 &  \\ 
LMC101.2.9556  & 5\uph20\upm30\zdot\ups53 & $-68\arcd46\arcm32\zdot\arcs3$ & $25.68$  & 0.11 & 0.27 & $106.06$  & 0.11 & 0.34 & 109.12 & 0.00  & 0.00 & 14.359 & 0.016 & 1.966 & 0.018 &  \\ 
LMC100.2.19195 & 5\uph20\upm43\zdot\ups03 & $-69\arcd22\arcm46\zdot\arcs5$ & $30.73$  & 0.16 & 0.27 & $117.17$  & 0.16 & 0.28 & 121.13 & 9.90  & 1.60 & 16.481 & 0.011 & 2.639 & 0.041 &  \\ 
LMC106.1.17195 & 5\uph21\upm26\zdot\ups11 & $-71\arcd49\arcm29\zdot\arcs6$ & $107.78$ & 0.87 & 0.32 & $-52.27$  & 0.87 & 0.37 & 119.79 & 0.00  & 0.00 & 19.427 & 0.098 & 2.565 & 0.367 &  \\ 
\ldots\\
\hline\\
\multicolumn{16}{p{18.5cm}}{Luminosity class is given as either WD, WDcand or subdwarf. Main sequence stars are not indicated. One can find three pairs of CPM binaries in the presented list.}
}
\end{landscape}

The star with the highest detected parallax found is LMC194.6.41 with $\pi
=91.3\pm1.6$~mas (\ie 8.4 times smaller than Proxima Centauri). It is a
main sequence star of spectral type $\approx$M5. The highest value of
proper motion found was $\mu=722.19\pm0.74$~mas/yr for LMC198.4.97 (\ie
14.3 times smaller than Barnard's star). Fig.~4 presents equatorial
coordinates of LMC194.6.41 and another exemplary star.
\begin{figure}[htb]
\centerline{\includegraphics[width=.9\textwidth]{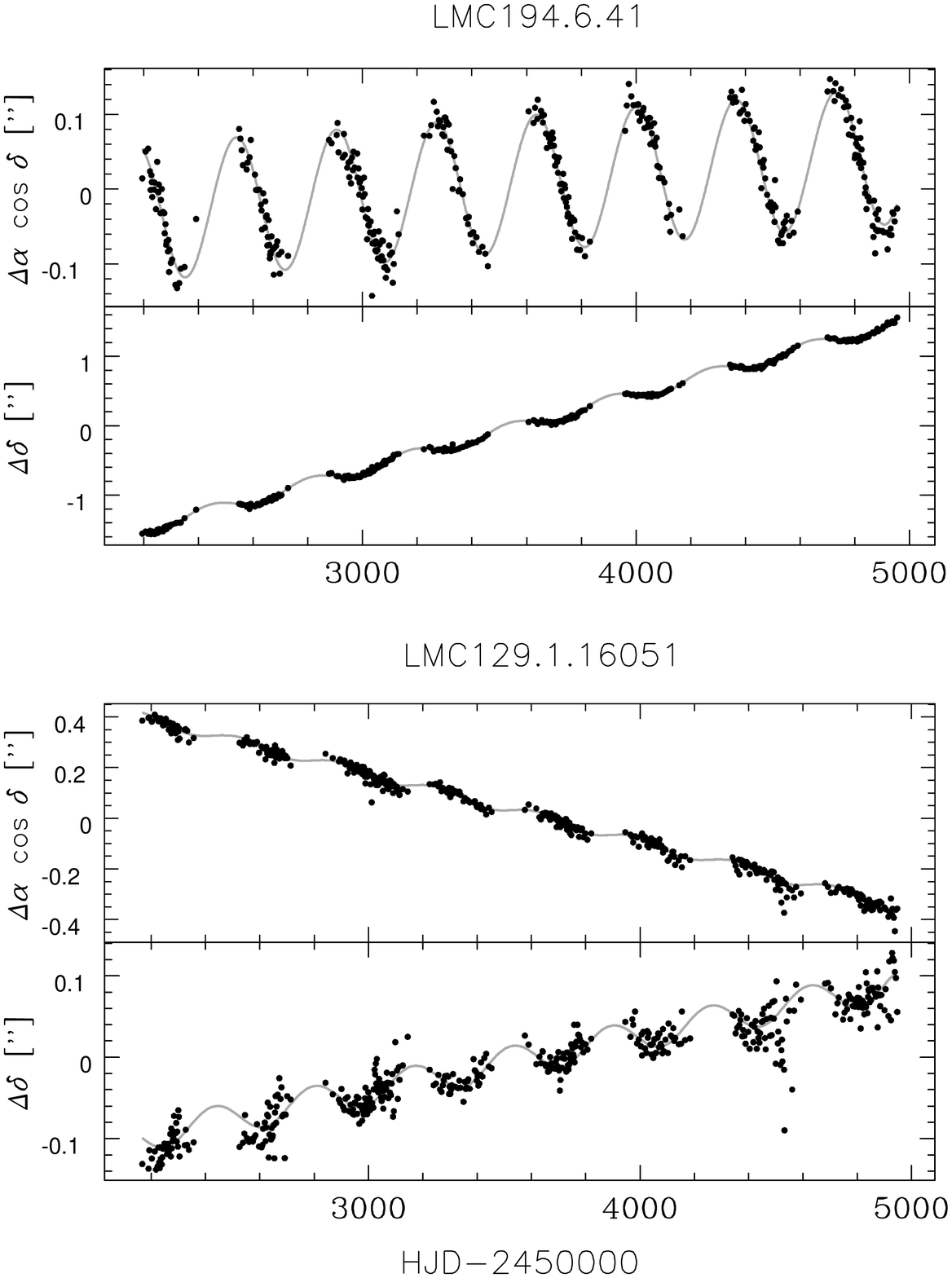}}
\FigCap{Equatorial coordinates as a function of time for two exemplary 
stars with model fits. The differential refraction effect was subtracted
from the data. The model parameters $\left(\mu_{\alpha\star},\mu_\delta,
\pi,r\right)$ are (10.1~mas/yr, 394.6~mas/yr, 91.3~mas, 11.3~mas) for 
LMC194.6.41 and ($-98.2$~mas/yr, 24.8~mas/yr, 19.1~mas, 20.2~mas) for
LMC129.1.16051.}
\end{figure}

\subsection{Completeness}
As was stated earlier we independently found 99\% of unique sources from
Alcock \etal (2001) and Soszyñski \etal (2002) catalogs with $\mu>95$~mas.
The number of objects in our catalog which are in the sky area covered by
OGLE-II and MACHO are 79 and 271, respectively, while these projects found,
respectively, 62 and 26 of these objects. Another estimate of the catalog
completeness may be performed internally. As mentioned previously some
stars are placed on overlapping parts of the frames and we should found
such stars independently. For our 551 stars in the catalog we have 584
useful (\ie with sufficient number of measurements) identifiers in OGLE-III
data -- 33 stars were present in adjacent subfields.  All of them were
present on our candidate HPM stars list. It suggests high completeness of
the catalog.

The SPM4 catalog (Girard \etal 2011) lists 7786 HPM objects located in the
MCs OGLE-III fields with {\it V}-band brightness similar to our stars \ie
between 14~mag and 21~mag. To check reliability of their findings we
randomly selected a sample of a hundred stars and retrieved time-series
astrometry from the database of raw {\sc DoPhot} results within a 3\arcs
radius circle.  The plots of $\alpha\cos\delta$ {\it vs.} $\delta$
coordinates with the color-coded epoch of measurements were examined in
detail. Only eight of these plots showed clearly moving objects. Two of
them are in our catalog and for other six the inferred proper motions are
below 50~mas/yr. We conclude that the reliability of SPM4 catalog in MCs
sky area is not sufficient to compare it with our list.

Fig.~5 shows sky projection of the HPM stars found. No obvious correlation
between the number of background stars and the number of HPM stars can be
seen, though, some regions, like Tarantula Nebula (field LMC175) and the
center of the SMC (field SMC100) show smaller number of HPM stars. We used
the Besan{\c c}on Milky Way model (Robin \etal 2003) to find the expected
number of HPM stars in the {\it I}-band brightness range
$12.8\div20.2$~mag. For the center of the LMC the model estimate was 4.3
HPM stars per area equivalent to the OGLE-III field. For the SMC the number
was 4.7. If compared to the observed numbers: 3.4 and 4.6 for the LMC and
SMC, respectively, this gave estimated completeness of 79\% and 98\%. The
biggest difference which plausibly increases the completeness in the SMC
fields as compared to the LMC fields was the larger number of epochs in the
SMC fields, but we do not think this could so much influence the
completeness. No other effect which may result in such a discrepancy was
found except possible small structures in the local number density of stars
which are not well characterized in the Besan{\c c}on model.

\begin{figure}[htb]
\centerline{\includegraphics[width=\textwidth]{fig5a.ps}}
\bigskip
\centerline{\includegraphics[width=.544\textwidth]{fig5b.ps}}
\vskip6pt
\FigCap{Positions of HPM stars found (black dots) in the foreground of 
the LMC ({\it upper panel}) and SMC ({\it lower panel}). Squares mark the
borders of OGLE-III fields with number given inside. The background images
originate from the ASAS survey (Pojmañski 1997).}
\end{figure}

\subsection{Comparison with OGLE-II and MACHO Catalogs}
While comparing the derived proper motions one has to keep in mind that if
the results in the two catalogs are based on images obtained a few years
apart, the blending with different background stars can be the main cause
of differences in the results. The blending is more probable in the dense
stellar regions and indeed we see the biggest differences for stars located
in the densest fields. Our catalog has 62 (26) common records with OGLE-II
(MACHO) catalog. After removing the most outlying results, we were left
with the {\it rms} of proper motion per coordinate differences equal to
3.3~mas/yr for OGLE-II and 5.9~mas/yr for MACHO catalog. Using OGLE-II and
our error bars we estimated the expected {\it rms} of 1.7~mas/yr. For MACHO
proper motions Alcock \etal (2001) only gave typical uncertainty of roughly
3.5~mas/yr per coordinate. There are eleven stars for which both we and
Soszyñski \etal (2002) gave estimates of parallax. Our results are on
average $2.5\pm1.0$~mas greater and the average uncertainty of our
measurements is 1.6~mas.

\subsection{Color--Magnitude Diagram}
For stars with parallax measurements we calculated the absolute magnitudes
in the {\it I}-band ($M_I$). They are shown as a function of $V-I$ color in
Fig.~6. The error-bars in $M_I$ include uncertainties of the {\it I}-band
photometry and parallax measurements. The points close to the lower right
corner have large $V-I$ error-bars because they are close to the detection
limit in the {\it V}-band. One can clearly see the main sequence stars most
of which have $(V-I)>2$~mag. The bluest objects (light blue triangles in
Fig.~6) are WDs which well separate from the main sequence stars. Between
these two groups there is a dozen or so stars which may be candidate
subdwarfs, although, all of them except one object (SMC110.5.999999;
$(V-I)=1.51$~mag, $M_I=12.1$~mag) have poorly known absolute magnitudes and
it is hard to select a clear sample of subdwarfs without additional
observations. Altogether 365 points are shown in Fig.~6; 20 objects were
rejected because their $V-I$ color is not known. Among 385 stars with
parallax measured the one with the faintest absolute brightness is
SMC108.6.8038 ($M_I=18.08\pm0.35$~mag).
\begin{figure}[htb]
\centerline{\includegraphics[width=.7\textwidth]{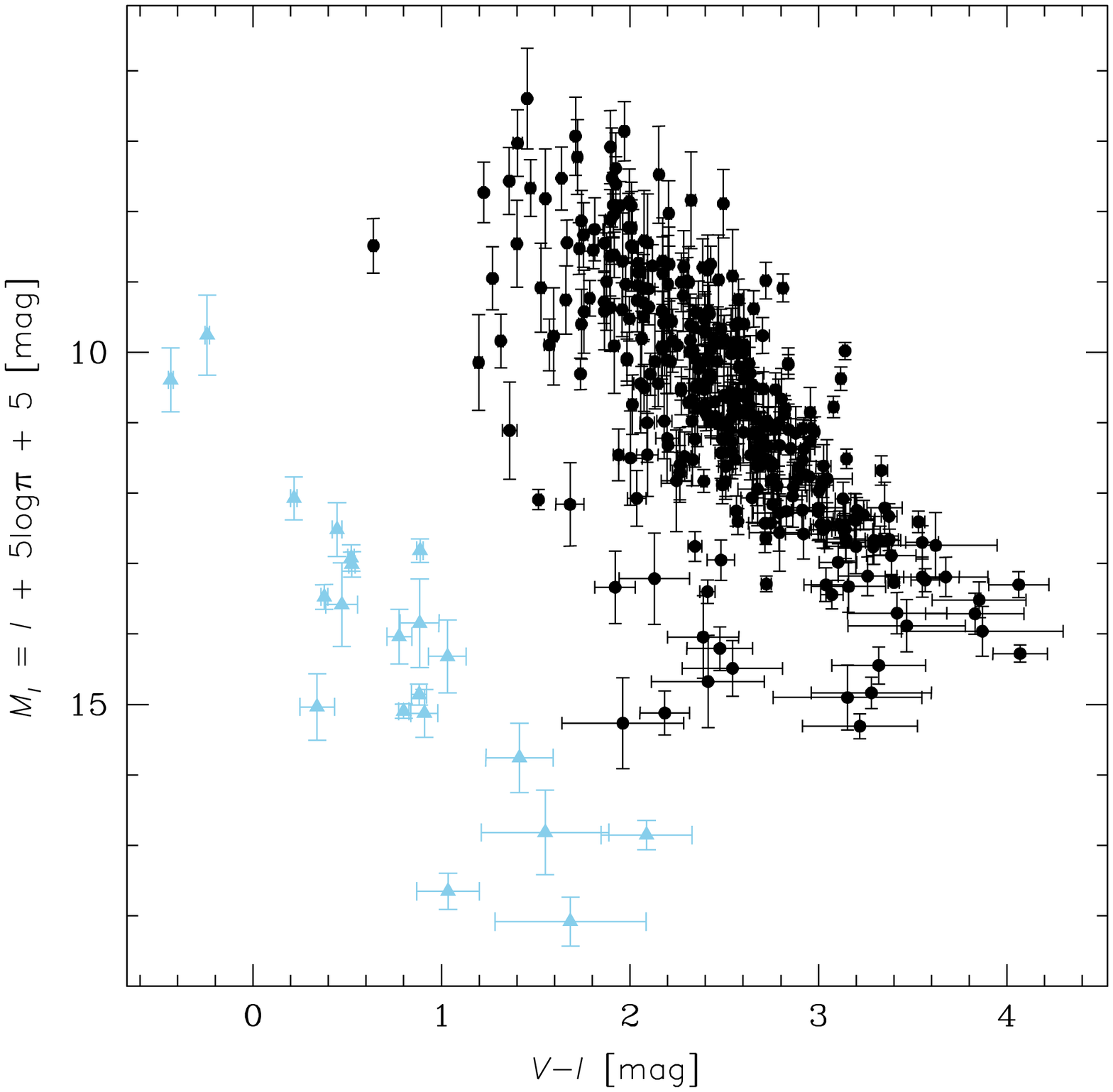}}
\FigCap{Color--absolute magnitude diagram for stars with measured
parallax. WDs are marked with light blue triangles.}
\centerline{\includegraphics[width=.7\textwidth]{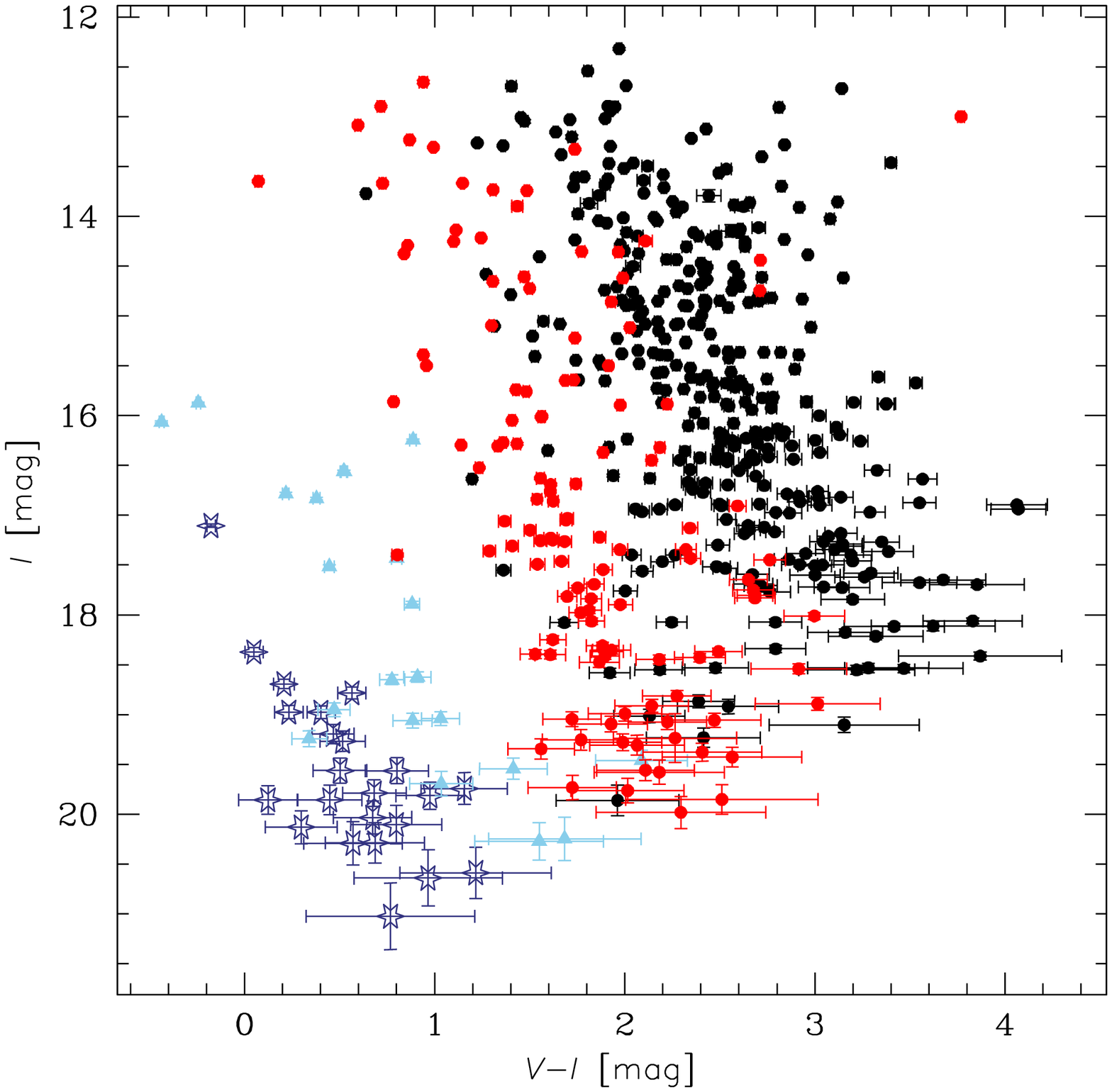}}
\FigCap{Color--magnitude diagram. Black and light blue symbols show the 
same stars as in Fig.~6. Dark blue symbols indicate candidate WDs and red
dots indicate all other stars without parallax measured.} 
\end{figure}

Fig.~7 shows the CMD for all HPM stars with color information. The stars
with parallax measured are shown using the same symbols as in Fig.~6. The
stars without parallax measured are shown using dark blue symbols, if they
are fainter or bluer than the locus of WDs from Fig.~6, and red dots
otherwise. These objects (dark blue symbols) are classified as WD
candidates. The red points indicate stars which most probably are main
sequence stars which are farther away than the stars marked with black
points.

\subsection{Common Proper Motion Binaries}
The list of confirmed stars with $\mu>95$~mas/yr was cross-matched with
itself and the pairs of stars located less than 5\arcm from each other were
checked if the values and the directions of their proper motions are
similar. Table~2 presents the list of 27 CPM binaries in which at least one
component had $\mu>100$~mas/yr. Six of these binaries are uncertain. The
separation of the components ranges from 0\zdot\arcs7 to 144\zdot\arcs3.
Large difference in brightness between components of CPM binary suggests
the fainter object may have very faint absolute brightness. We indicated
the difference in the {\it I}-band brightness, if it was larger than
3.5~mag.

\renewcommand{\arraystretch}{0.95}
\MakeTableee{llccl}{12.5cm}{Common proper motion binaries}
{\hline
\multicolumn{1}{c}{ID1}\uprule &
\multicolumn{1}{c}{ID2} &
\multicolumn{1}{c}{angular dist.} &
\multicolumn{1}{c}{$\mu$} &
\multicolumn{1}{c}{comments} \\
\dorule &
&
\multicolumn{1}{c}{$\left[\arcs\right]$} &
\multicolumn{1}{c}{$[{\rm mas/yr}]$} &
\\
\hline
\uprule
LMC162.4.41184 & LMC162.4.41229 & ~~~~0.7 & 151 &  \\ 
LMC102.2.30    & LMC102.2.37    & ~~~~0.9 & 141 &  \\ 
LMC130.3.25    & LMC130.3.4226  & ~~~~1.2 & 107 &  \\ 
LMC120.6.17556 & LMC120.6.17583 & ~~~~1.4 & 168 &  \\ 
LMC179.4.38816 & LMC179.4.40244 & ~~~~1.4 & 111 & uncertain; $\Delta I=4.7$~mag\\ 
SMC125.7.29221 & SMC125.7.29234 & ~~~~1.7 & 124 &  \\ 
LMC174.8.25010 & LMC174.8.31202 & ~~~~2.9 & 102 & $\Delta I=3.6$~mag\\ 
SMC111.8.20419 & SMC111.8.20441 & ~~~~3.7 & 102 &  \\ 
LMC146.8.30    & LMC146.8.31    & ~~~~3.7 & 106 &  \\ 
LMC155.1.4867  & LMC155.1.5999  & ~~~~3.7 & 100 &  \\ 
SMC133.4.124   & SMC133.4.3322  & ~~~~3.9 & 106 & uncertain \\ 
LMC166.5.30    & LMC166.5.499   & ~~~~4.1 & 205 &  \\ 
LMC161.1.38    & LMC161.1.5     & ~~~~6.5 & 113 &  \\ 
LMC126.1.100   & LMC126.1.9     & ~~~~6.6 & 127 &  \\ 
LMC176.1.34425 & LMC176.1.34431 & ~~~~6.7 & 111 &  \\ 
SMC139.7.1569  & SMC139.7.1570  & ~~~~8.2 & 103 &  \\
LMC119.5.40603 & LMC119.5.40836 & ~~~~8.2 & 144 &  \\ 
SMC115.5.12    & SMC115.5.319   & ~~~~9.1 & 315 & $\Delta I=3.8$~mag\\ 
LMC130.6.186   & LMC130.6.29    &  ~~10.6 & 151 & uncertain \\ 
LMC186.4.38205 & LMC186.4.38233 &  ~~16.0 & 181 & uncertain \\ 
SMC110.8.23710 & SMC110.8.23860 &  ~~19.0 & 180 &  \\ 
LMC103.3.2     & LMC103.6.38675 &  ~~58.3 & 126 &  \\ 
SMC114.6.11666 & SMC114.7.14379 &  ~~65.2 & 113 &  \\ 
SMC125.6.4503  & SMC125.6.8973  &  ~~86.0 & 126 & uncertain \\ 
LMC102.7.22769 & LMC102.7.22886 &  ~~92.0 & 119 &  \\ 
LMC211.1.10    & LMC211.8.3926  &   114.3 & 208 &  \\ 
LMC106.1.14252 & LMC106.1.17195 &   147.3 & 115 & uncertain \\ 
\hline
\noalign{\vskip9pt}
\multicolumn{5}{p{12.5cm}}{The brightness difference $\Delta I$ is given if larger than 3.5~mag.}
}

\Section{Summary} 
The presented catalog contains 549 stars with $\mu>100$~mas/yr observed in
the direction to the Magellanic Clouds, which are dense stellar regions.
The highest proper motion found is $722.19\pm0.74$~mas/yr. For 70\% of
these objects parallaxes were measured with significance greater than
3$\sigma$. The closest object has $\pi=91.3\pm1.6$~mas. The completeness of
the catalog is higher by at least 27\% than the previous ones investigating
the same sky area. Altogether 44 objects are marked as WDs or candidate
WDs.The search for CPM binaries resulted in 27 pairs. The catalog may be
useful for estimation of the local number density of intrinsically faint
objects. The follow-up observations in the infrared or cross-matching with
existing catalogs may result in a candidate brown dwarfs in the close
proximity of our HPM. The MCs host many different types of astrophysically
important objects and are the nearest such galaxies, thus follow-up
observations may be performed when other targets are observed. Jointly with
other catalogs this may extend further our knowledge of the Galaxy
dynamics, the motion of the Sun against nearby stars and wide binary
systems.

The uncertainties of our proper motions are around 0.5~mas/yr. Our results
were compared to the other catalogs and the major contribution to
differences found came from blending of the HPM stars with different
objects in different surveys. The catalog of stars with $\mu<100$~mas/yr
toward the MCs will be presented in the forthcoming paper. A continuation
of our efforts will be a calculation of proper motions in the OGLE-III
Galactic Bulge fields.

\Acknow{Authors thank S. Koz³owski and P. Pietrukowicz for comments on 
manuscript. This work was supported by MNiSW grant N-N203-512538. RP is
supported through Polish Science Foundation START program. The OGLE project
has received funding from the European Research Council under the European
Community's Seventh Framework Programme (FP7/2007-2013)/ERC grant agreement
No. 246678.}

\end{document}